\documentclass{IEEEtran}
\usepackage[utf8]{inputenc}

\title{Learning-based Bounded Synthesis for Semi-MDPs with LTL Specifications}
\author{Ryohei Oura }
\date{October 2021}

\usepackage{amsfonts,amssymb}
\usepackage{nccmath}
\usepackage{bm}
\usepackage{listings}
\usepackage{float}
\usepackage{comment}
\usepackage{algorithm}
\usepackage{algorithmic}
\usepackage{varwidth}
\usepackage{enumitem}
\usepackage{diagbox}
\usepackage{graphicx}
\usepackage{subfigure}
\usepackage{dsfont}
\usepackage{caption}

\usepackage{lipsum}
\allowdisplaybreaks[1]

\newtheorem{theorem}{Theorem}

\newtheorem{assumption}{Assumption}

\newtheorem{definition}{Definition}
\newtheorem{notations}{Notations}

\newtheorem{runexample}{Running example}

\newtheorem{remark}{Remark}

\DeclareMathOperator*{\argmin}{arg\,min}
\DeclareMathOperator*{\argmax}{arg\,max}

\author{Ryohei~Oura and
        ~Toshimitsu~Ushio,~\IEEEmembership{Member,~IEEE,}
\thanks{
This work was partially supported by JST-ERATO HASUO Project Grant Number JPMJER1603, Japan and JST CREST Grant Number JPMJCR2012, Japan.
}
\thanks{
The authors are with the Graduate School of Engineering Science, Osaka University, Toyonaka 560-8531, Japan (e-mail: r-oura@hopf.sys.es.osaka-u.ac.jp; ushio@sys.es.osaka-u.ac.jp).
}
}

\begin{document}

\maketitle

\begin{abstract}
This letter proposes a learning-based bounded synthesis for a semi-Markov decision process (SMDP) with a linear temporal logic (LTL) specification. In the product of the SMDP and the deterministic $K$-co-B\"uchi automaton (d$K$cBA) converted from the LTL specification, we learn both the winning region of satisfying the LTL specification and the dynamics therein based on reinforcement learning and Bayesian inference. Then, we synthesize an optimal policy satisfying the following two conditions. (1) It maximizes the probability of reaching the wining region. (2) It minimizes a long-term risk for the dwell time within the winning region. The minimization of the long-term risk is done based on the estimated dynamics and a value iteration. We show that, if the discount factor is sufficiently close to one, the synthesized policy converges to the optimal policy as the number of the data obtained by the exploration goes to the infinity.

\end{abstract}

\begin{IEEEkeywords}
Bounded Synthesis, Linear Temporal Logic, Reinforcement Learning, Bayesian inference, Semi-Markov Decision Process.
\end{IEEEkeywords}

\section{Introduction}
Linear temporal logic (LTL) is used to describe a complex control specification \cite{BK2008}. In recent years, several automata-guided reinforcement learning (RL) have been investigated to synthesize an optimal controller for an unknown controlled stochastic system, e.g., Markov decision process (MDP), under an LTL specification \cite{SKCSS2014, BWZP2020, HPSSTW2019, OST2020}. They constructed the product MDP from the MDP and a Rabin or B\"uchi automaton converted from the LTL specification. An optimal controller was learned on the product MDP by maximizing the expected return with a reward assignments according to the acceptance condition. In \cite{LVB2017, LSYB2019}, truncated LTL specifications were used and rewards were assigned according to the converted automaton. Bounded synthesis is an alternative synthesis approach to avoid the state-complexity of Rabin automata \cite{FJR2011}. A bounded synthesis and reinforcement learning-based optimal controller synthesis for MDPs has been proposed \cite{OUS2021}.



On the other hand, many systems have intrinsic stochasticity of a time interval between state transitions, namely dwell time, as well as that of state transition. The stochasticity of the dwell time may degrade a quantitative performance of the controller that satisfies the LTL specification.  For example, a round trip time in patrol  of a museum by a mobile robot depends on dwell times in exhibition rooms. So, in practice, it is important to take a \textit{risk} for the dwell time, e.g., the variance of the dwell time, into account to achieve a long-term task. 
A semi-Markov decision process (SMDP) is leveraged as a model of a controlled stochastic system to represent the stochasticity of the dwell time and that of the state transition \cite{RS2013}. Several reinforcement learning methods for SMDPs have been advocated \cite{BD1995, SS2021}. They mainly aimed at maximizing an accumulative rewards in the continuous time. 
To the best of our knowledge, however, reinforcement learning of a control policy satisfying an LTL specification under the stochastic dwell time has not been addressed.
Several model-based RL methods for LTL specifications and cost minimization have been addressed \cite{WBL2017, WT2016, FT2014}. However, they only considered the (almost) sure satisfaction of the LTL specifications and applied a model-based RL to all states.


In this letter, we propose a bounded synthesis and learning-based method for the optimal policy synthesis that minimizes a long-term risk for the dwell time under the satisfaction of a given LTL specification with the maximal probability. First, we represent the unknown controlled system by an SMDP and convert the given LTL specification into a d$K$cBA. Then, for the product SMDP from them, we synthesize an optimal policy satisfying the following two conditions. (1) It maximizes the probability of reaching the winning region. (2) It minimizes the risk for the dwell time within the winning region.
The rest of this letter is organized as follows. Section \ref{Pre} reviews an SMDP, LTL, and $\omega$-automata. Section \ref{ProbSta} formulates a control problem given by an LTL specification with a risk by the dwell time. Section \ref{Method} proposes a bounded synthesis and learning-based method for an optimal policy synthesis. 

\begin{notations}
$\mathbb{N}$ is the set of positive integers. $\mathbb{N}_0$ is the set of non-negative integers. $\mathbb{R}$ is the set of real numbers. $\mathbb{R}_{\geq 0}$ is the set of non-negative real numbers. For a set $A$, we denote its cardinality by $|A|$ and the set of probability distributions over $A$ by $\mathcal{D}(A)$. For two sets $A$ and $B$, denoted by $A^B$ is the set of mappings from $B$ to $A$. 
We denote the empty string by $\varepsilon$.
\end{notations}

\section{Preliminaries}
\label{Pre}

\subsection{Semi-Markov decision process}

A semi-Markov decision process (SMDP) is a continuous-time dynamic system \cite{RS2013, BD1995, SS2021} modeled by a tuple $\mathcal{SM} = (S, A, T, D, s^I, AP, L)$, where $S$ is the finite set of sates, $A$ is the finite set of actions, $T : S \times S \times A \to [0,1]$ is the state transition probability, $D : S \times A \times S \to \mathcal{D}(\mathbb{R}_{\geq 0})$ is the dwell time probability function, $s^I \in S$ is the initial state, $AP$ is the finite set of atomic propositions, and $L : S \to 2^{AP}$ is the labeling function \cite{BK2008}. Let $A(s) = \{ a \in A \;|\; \exists s' \in S \mbox{ s.t. } T(s' \;|\; s,a) > 0 \}$. Note that $\sum_{s' \in S}T(s' \;|\; s,a) = 1$ for any $s \in S$ and any $a \in A(s)$. We sometimes denote $D(\cdot \;|\; s,a,s')$ instead of $D(s,a,s')(\cdot)$ for any $(s,a,s') \in S \times A \times S$.

\begin{remark}
The semi-Markov kernel $P : S \times \mathcal{B}(\mathbb{R}_{\geq 0}) \times S \times A \to [0,1]$, where $\mathcal{B}(\mathbb{R}_{\geq 0})$ is the Borel set over $\mathbb{R}_{\geq 0}$, is defined as $P(s', \tau \;|\; s, a) = T(s' \;|\; s,a) D(\tau \;|\; s,a,s')$ for any $(s,a,s') \in S \times A \times S$ and any $\tau \in \mathcal{B}(\mathbb{R}_{\geq 0})$.
\end{remark}


In the SMDP $\mathcal{SM}$, an infinite (time-abstract) path starting from a state $s_0 \in S$ is defined as a sequence $\rho\ =\ s_0a_0s_1 \ldots\ \in S (A S)^{\omega}$ such that $T(s_{i+1}|s_i, a_i) > 0$ for any $ i \in \mathbb{N}_0$. A finite path is a finite sequence in $S (A S)^{\ast}$. In addition, we sometimes represent $\rho$ as $\rho^I$ to emphasize that $\rho$ starts from $s_0 = s^I$.
For a path $\rho\ =\ s_0a_0s_1 \ldots$, we define the corresponding labeled path $L(\rho)\ =\ L(s_0)L(s_1) \ldots \in (2^{AP})^{\omega}$. 
$InfPath^{SM}\ ( \text{resp., }FinPath^{SM})$ is defined as the set of infinite (resp., finite) paths starting from $s_0=s^I$ in the SMDP $\mathcal{SM}$. 
For each finite path $\rho$, $last(\rho)$ denotes its last state.

Let $InfPath^{\mathcal{SM}}_{\pi}$ (resp., $FinPath^{\mathcal{SM}}_{\pi}$) be the set of infinite (resp., finite) paths starting from $s_0=s^I$ in the SMDP $\mathcal{SM}$ under a policy $\pi$. The time-abstract behavior of the SMDP $\mathcal{M}$ under a policy $\pi$ is defined on a probability space $(InfPath^{\mathcal{SM}}_{\pi}, \mathcal{F}_{InfPath^{\mathcal{SM}}_{\pi}}, Pr^{\mathcal{SM}}_{\pi})$ \cite{BK2008}. 

\subsection{Linear Temporal Logic and $\omega$-automata}
A linear temporal logic (LTL) formula is used to describe a complex control specification. See \cite{BK2008} for the detail of LTL.

For an LTL formula $\varphi$ and an infinite path $\rho = s_0a_0s_1 \ldots$ of the SMDP $ \mathcal{SM} $, we denote the satisfaction relation as $\mathcal{SM},\rho \models \varphi$ and sometimes we omit $\mathcal{SM}$.

An $\omega$-automaton is a tuple $A = (X,\Sigma,\delta,x^I,Acc)$, where $X$ is the finite set of states, $\Sigma$ is the input alphabet, $\delta \subseteq  X\times \Sigma \times X$ is the set of transitions, $x^I \in X$ is the initial state, and $Acc \subseteq X$ is the accepting set.

We say that $A$ is deterministic if $|\; \{ x^\prime \in X \;|\; (x, \sigma, x^\prime) \in \delta \} \;| \leq 1$ for any $x \in X$ and any $\sigma \in \Sigma$. $A$ is complete if $ \{ x^\prime \in X \;|\; (x, \sigma, x^\prime) \in \delta \} \neq \emptyset$ for any $x \in X$ and any $\sigma \in \Sigma$.

An infinite sequence $w \in \Sigma^{\omega}$ is called a word. An infinite sequence $r = x_0\sigma_0x_1 \ldots \in X (\Sigma X)^{\omega}$ is called a run on $A$ generated by a word $w = \sigma_0 \sigma_1 \ldots \Sigma^\omega$ if $(x_i, \sigma_{i}, x_{i+1}) \in \delta\ $ for any $ i\in \mathbb{N}_0$.  For an $\omega$-automaton $A$ and a word $w$, we denote by $Runs(w; A)$ the set of runs on $A$ generated by $w$. Moreover, for a state $x$ and a run $r$, we denote by $Visits(x; r)$ the number of times $r$ visits $x$. An $\omega$-automaton $A$ is called a universal co-B\"uchi automaton (cBA) (\textit{resp.}\ $K$-co-B\"uchi automaton ($K$cBA) )  if (\ref{eq:cBA}) (\textit{resp.} (\ref{eq:KcBA}) ) holds. 

\begin{align}
\forall r \in Runs(w;A), \forall x \in Acc. Visits(x; r)<\infty.
\label{eq:cBA}
\end{align}

\begin{align}
\forall r \in Runs(w;A). \sum_{x \in Acc} Visits(x; r) \leq K.
\label{eq:KcBA}
\end{align}

We denote the sets of words accepted by a cBA and a $K$cBA by ${\mathcal L}_c(A)$ and  ${\mathcal L}_{c,K}(A)$, respectively.
Moreover, the co-B\"uchi automaton (cBA) and the $K$-co-B\"uchi automaton ($K$cBA) are denoted by $B$ and $(B,K)$, respectively, to make clear which acceptance condition is adopted.


We construct a determinization of $K$cBA (d$K$cBA) $det(B,K) = (\mathcal{F}, \Sigma, \Delta, F^I, Acc_{d})$ from the $K$cBA $(B,K) = (X, \Sigma,\delta,x^I,Acc)$ by a subset construction scheme combined with the counter for visits to $Acc$. 
See \cite{FJR2011} for the detailed construction method.
For any $K \in \mathbb{N}_0$, $det(B,K)$ is deterministic and complete, and satisfies
\begin{align}
    \mathcal{L}_{c,K}(B) = \mathcal{L}_{c,0}(det(B,K)).
    \label{words_KdK}
\end{align}

It is known that, for any LTL formula $ \varphi $, there exists a cBA $B_\varphi$ that accepts all words satisfying $\varphi$ \cite{FJR2011}, where the alphabet is given by $ \Sigma = 2^{AP} $.

Let $A = (X,\Sigma,\delta,x^I,Acc)$ be an $\omega$-automaton. For the subset of states $X_{sub} \subseteq X$, we call $X_{sub}$ a sink set if there is no outgoing transition from $X_{sub}$ to $X \setminus X_{sub}$, namely, $ \{ (x, \sigma, x^\prime) \in \delta \;|\; x \in X_{sub}, \sigma \in \Sigma, x^\prime \in X \setminus X_{sub} \} = \emptyset $.

Note that the set of accepting sates $Acc_d$ can be constructed as a sink set for any d$K$cBA $det(B,K)$. This is because, for any run $r \in X(\Sigma X)^\omega$, once $r$ enters $Acc_d$, it never satisfies the acceptance condition of $det(B,K)$. Without loss of generality, $Acc_d$ can be constructed as a singleton.

For any policy $\pi$ and any state $s \in S$, the probability of paths starting from $s$ satisfies an LTL formula $\varphi$ on the SMDP $\mathcal{SM}$ under $\pi$ is defined as follows.

\begin{align*}
Pr^{\mathcal{SM}}_{\pi}(s \models \varphi) := Pr^{\mathcal{SM}}_{\pi}(\{ \rho \! \in \! InfPath^{\mathcal{SM}}_{\pi}(s) \;|\; \rho \! \models \varphi\}).
\end{align*}
Similarly, we define the probability of paths starting from the initial state $s$ and the initial action $a$ satisfies $\varphi$ as $Pr^{\mathcal{SM}}_{\pi}(s,a \models \varphi)$.
We call $Pr^{\mathcal{SM}}_{\pi}(s^I \models \varphi)$ \textit{the satisfaction probability} of $\varphi$ on $\mathcal{SM}$ under $\pi$.

\subsection{Bayesian inference}
To estimate the state transition and the dwell time probability, we use the following Bayesian inference procedure \cite{B2006}.

Let $\Theta$ be the metric space of parameters and $O$ be the set of observations. For $\mathcal{O} \subseteq O$ and $\bm{\theta} \in \Theta$, we first introduce a parametric model $p(\mathcal{O} | \bm{\theta})$ and choose a prior $p(\bm{\theta})$. Next, we observe data $\mathcal{O}_n = \{ \bm{o}_0, \bm{o}_1, \ldots, \bm{o}_n \}$ from the system. We then calculate the posterior
\begin{align}
    p( \bm{\theta} | \mathcal{O}_n ) \propto p(\mathcal{O}_n | \bm{\theta}) p(\bm{\theta}).
    \label{bayesrule}
\end{align}

Doob's theorem guarantees that the posterior probability concentrates in the neighborhood of the true parameter under mild conditions as $n \to \infty$ \cite{M2018}, namely the estimated parameter converges to the true parameter in probability.

\begin{assumption}
    Assume the following conditions.
    \begin{enumerate}
        \item $\bm{\theta} \mapsto p(\cdot \;|\; \bm{\theta})$ is measurable.
        \item $\bm{\theta} \neq \bm{\theta}^\prime$ implies $p(\cdot \;|\; \bm{\theta}) \neq p(\cdot \;|\; \bm{\theta}')$.
    \end{enumerate}
    \label{assum_Doob}
\end{assumption}

\begin{theorem}[\cite{M2018}]
    Under Assumption \ref{assum_Doob}, there exists $\Theta_1 \subseteq \Theta$  with $p(\Theta_1) = 1$ and, for any $\bm{\theta}_1 \in \Theta_1$, if $\mathcal{O}_n = \{\bm{o}_1, \ldots, \bm{o}_n \} \sim p(\cdot \;|\; \bm{\theta}_1) \mbox{ i.i.d.}$ then, for any neighborhood $B$ of $\bm{\theta}_1$, we have $ \lim_{n\to \infty} p(\bm{\theta} \in B \;|\; \mathcal{O}_n) = 1 \mbox{ a.s. } [ p(\cdot \;|\; \bm{\theta}_1) ].$
    \label{th_Doob}
\end{theorem}
 
\section{Problem Statement}
\label{ProbSta}
We consider the SMDP $\mathcal{SM}$ with unknown $T$ and $D$ and the LTL specification $\varphi$.
We aim to synthesize a control policy that minimizes the long-term risk for the dwell time of $\mathcal{SM}$ under the achievement of $\varphi$ with the  maximal probability.

To determine the satisfaction of $\varphi$ for $\mathcal{SM}$, we introduce the product SMDP $\mathcal{SM}^\otimes$ of the SMDP $\mathcal{SM} $$=$$ (S, A, T, D, s^I, AP, L)$ and the d$K$cBA $det(B_\varphi,K) $$=$$ (\mathcal{F}, \Sigma,\Delta,F^I,Acc_d)$. $\mathcal{SM}^\otimes$ is given by the tuple $(S^{\otimes},  A^{\otimes}, T^{\otimes}, D^\otimes, s^{\otimes I}, Acc^{\otimes})$, where
$S^{\otimes} = S \times F$ is the finite set of states; $s^{\otimes I} = (s^{I},\hat{F}^{I})$ is the initial states, where $(F^I, L(s^I), \hat{F}^I) \in \Delta$;
$A^\otimes$$=$$A$ and $\mathcal{A}^\otimes(s^\otimes)$$=$$\mathcal{A}(s)$ for each $s^\otimes$$=$$(s,F) \in S^\otimes$; $T^{\otimes} : S^{\otimes} \times S^{\otimes} \times A^{\otimes} \to [0,1]$ is the transition probability defined as, for any $s^\otimes = (s,F), s^{\otimes \prime} = (s',F') \in S^\otimes$ and any $a \in A(s)$,
\begin{align}
  &T^{\otimes}(s^{\otimes \prime} | s^{\otimes}, a) =
  &\left\{
  \begin{aligned}
    &T(s^{\prime} | s, a) &   &\text{if}\  (F, L(s), F^{\prime}) \in \Delta, \\
    &0 &   &\text{otherwise} ,
  \end{aligned}
  \right. 
  \label{Ttimes}
\end{align}
$D^\otimes : S^\otimes \times A \times S^\otimes \to \mathcal{\mathbb{R}}_{\geq 0})$ is the dwell time probability defined as, for any $(s^\otimes = (s,F), a, s^{\otimes \prime}=(s',F')) \in S^\otimes \times A \times S^\otimes$, $D^\otimes(s^\otimes, a, s^{\otimes \prime}) = D(s,a,s')$, where $s^\otimes = (s,F)$ and $s^{\otimes \prime} =(s', F')$, and $Acc^{\otimes} = S \times Acc_d$.
Note that $Acc^\otimes$ is non-empty since $Acc_d$ is non-empty.
In the following, the product SMDP $\mathcal{SM}^{\otimes}$ of a given SMDP and a d$K$cBA converted from an LTL formula $\varphi$ will be called a product SMDP associated with $\varphi$.
For any product SMDP $\mathcal{SM}^\otimes$, its acceptance condition is a safety condition since it is satisfied when any path generated on $\mathcal{SM}^\otimes$ under a policy always stays in $S^\otimes \setminus Acc^\otimes$.

For a subset $S_{sub}^\otimes$ of $S^\otimes$, we introduce an atomic proposition ``This state belongs to $S_{sub}^\otimes$'', which denotes $S_{sub}^\otimes$ by abuse of notation. Namely, we say that a state $s\in S^{\otimes}$ satisfies $S^\otimes_{sub}$ if $s \in S^\otimes_{sub}$. Then, the acceptance condition of $\mathcal{SM}^\otimes$ is represented by
\begin{align}
\varphi_B = \Box \neg Acc^\otimes.
\end{align}

A policy on the product SMDP $\mathcal{SM}^\otimes$ is defined as a mapping $\pi: S^\otimes \times A \rightarrow [0,1]$. A policy $\pi$ is {\it positional} if, for any $ s \in S^\otimes$ and any $ a \in \mathcal{A}^\otimes(s)$, there exists one action $ a' \in \mathcal{A}^\otimes(s)$ such that $\pi(s, a) = 1$ if $a=a^{\prime}$, and $\pi(s, a) = 0$ for any $ a \in \mathcal{A}^\otimes(s) $ with $a\neq a'$. We sometimes denote a positional policy $\pi$ as a mapping from $S^\otimes$ to $A$, that is, $\pi(s) = a$ for any $(s,a) \in S^\otimes \times A$ if $\pi(s,a)=1$.

In the following, we sometimes omit the superscript $\otimes$. We define the set of winning pairs $W_p \subseteq S^\otimes \times A$ and the winning region $W \subseteq S^\otimes$ as follows.
\begin{align*}
    &W = \{ s \in S^\otimes \;|\; \exists \pi \mbox{ s.t. } Pr^{\mathcal{SM}^\otimes}_{\pi}(s \models \varphi_B) = 1\}.\\
    &W_p = \{ (s,a) \in S^\otimes \times A \;|\; \exists \pi \mbox{ s.t. } Pr^{\mathcal{SM}^\otimes}_{\pi}(s,a \models \varphi_B) = 1\}.
\end{align*}

We define a risk factor for the dwell time $\mathsf{Risk} : S \times A \times S \to \mathbb{R}_{\geq 0}$ as follows. For any $(s,a,s')\in S \times A \times S$,
\begin{align}
    \mathsf{Risk}(s,a,s') = f(D(s,a,s')),
    \label{riskD}
\end{align}
where $f : \mathcal{D}(\mathbb{R}_{\geq 0}) \to \mathbb{R}$. As examples of $f$, we have $f_\alpha(d) = \inf \{ t \;|\; d(\tau > t) < \alpha \}$ with $\alpha \in [0,1]$ and $f_\lambda(d) = \mu + \lambda \sigma$ with $\lambda \in [0,1]$, where $\mu$ and $\sigma$ are the mean and the standard deviation of $d$, respectively.

 For a policy $\pi$ on the SMDP $\mathcal{SM}$ and a risk factor for the dwell time $\mathsf{Risk}$, we define the value function $V^{\mathsf{Risk}}_{\pi} : W \to \mathbb{R}$ for the risk factor as follows. For any $s \in W$,
  \begin{align*}
    V^{\mathsf{Risk}}_{\pi}(s)= \mathbb{E}_{\pi}[\sum_{n=0}^{\infty}\gamma^n_{\mathsf{r}} \mathsf{Risk}(s_n, a_n, s_{n+1})\;|\;s_0 = s],
  \end{align*}
where $\mathbb{E}_{\pi}$ denotes the expected value given that the action selection follows the policy $\pi$ from the state $s$ and $\gamma_{\mathsf{r}} \in [0,1)$ is a discount factor.
We define the action-value function $Q^{\mathsf{Risk}}_{\pi} : W_p \to \mathbb{R}$ as follows. For any $(s,a) \in W_p$, 
  \begin{align*}
    Q^{\mathsf{Risk}}_{\pi}(s,a)= \mathbb{E}_{\pi}[\sum_{n=0}^{\infty}\gamma^n_{\mathsf{r}} \mathsf{Risk}(s_n, a_n, s_{n+1})\;|\;s_0 = s, a_0=a]
  \end{align*}

Then, we aim at the synthesis of a positional policy $\pi^\ast : S^\otimes \to A$ on the product SMDP $\mathcal{SM}^\otimes$ that satisfies the following conditions:
\begin{enumerate}
    \item $\pi^\ast \in \Pi^\ast_{\varphi_B}(s^{\otimes I})$.
    \item $\pi^\ast \in \argmin_{\pi \in \Pi^\ast_{\varphi_B}(s^\otimes)} V^{\mathsf{Risk}}_{\pi}(s), \forall s^\otimes = (s,F) \in W$,
\end{enumerate}
where $\Pi^\ast_{\varphi_B}(s^\otimes) = \argmax_{\pi \in A^{S^\otimes}} Pr^{\mathcal{SM}^\otimes}_{\pi}(s^\otimes \models \varphi_B)$ for any $s^\otimes \in S^\otimes$.
We say that the above policy $\pi^\ast$ is optimal with respect to $\varphi_B$ and $\mathsf{Risk}$ or simply optimal. We denote the set of optimal policies as $\Pi^\ast_{\varphi_B \succ \mathsf{Risk}}$.

\begin{runexample}
We consider the surveillance problem of a mobile robot in a grid world depicted in Fig.\ \ref{fig:grid} as a running example. The set of states is $S = \{1,\ldots,5 \}^2$. The set of actions is $A = \{ UL, UR, DL, DR\}$. The black disk represents the robot and its initial state is the right corner $(5,5)$. The robot moves up and left with probabilities 0.5 under the actions $UL$. Under $UR$, it moves up and right with probabilities of 0.5, respectively. Likewise, $DL$ and $DR$ trigger each transition to down, left, and right with probabilities of 0.5.
Under an action, if the robot faces a wall, it makes the transition to a movable direction w.p.1. If the robot has no movable direction under an action, it stays in the same state. The blue and green areas represent the recharge point and the access point, which are labeled with $a$ and $b$. The red area represents the dangerous area, which is labeled with $c$. The dwell time probability is given by the exponential distribution $D(d \tau \;|\; s,a,s') = \lambda_{sas'} e^{-\lambda_{sas'} \tau} d \tau $ for each transition $(s,a,s')$. The parameter $\lambda_{sas'}$ is given by $10 \max \{ s_x-3, s_y-3 \}$, where $s = (s_x, s_y)$, which implies that the closer the robot is to the wall, the larger the tail of the dwell time distribution is.
The specification is given as ``Visit the blue and green areas infinitely often while avoiding the red area". The specification formally represented as
\begin{align}
\varphi_{ex} = \Box \diamondsuit a \land \Box \diamondsuit b \land \neg \Box c. 
\label{spec:ex}
\end{align}
The d$K$cBA with $K=20$ converted from $\varphi_{ex}$ is shown in Fig.\ \ref{fig:dKcBA}.
Intuitively, the robot is required to visit the blue and the green areas within $K=20$ time steps infinitely often.
The robot aims to satisfy $\varphi_B$ obtained from $\varphi_{ex}$ with the maximal probability and simultaneously minimize the accumulative risk of the dwell time. When we set $\mathsf{Risk}(s,a,s') = \lambda_{sas'}$, the robot has to move as near the state $(3,3)$ as possible.
\begin{figure}[htbp]
    \subfigure[]{
        \centering
        \includegraphics[clip, width = 3.9cm]{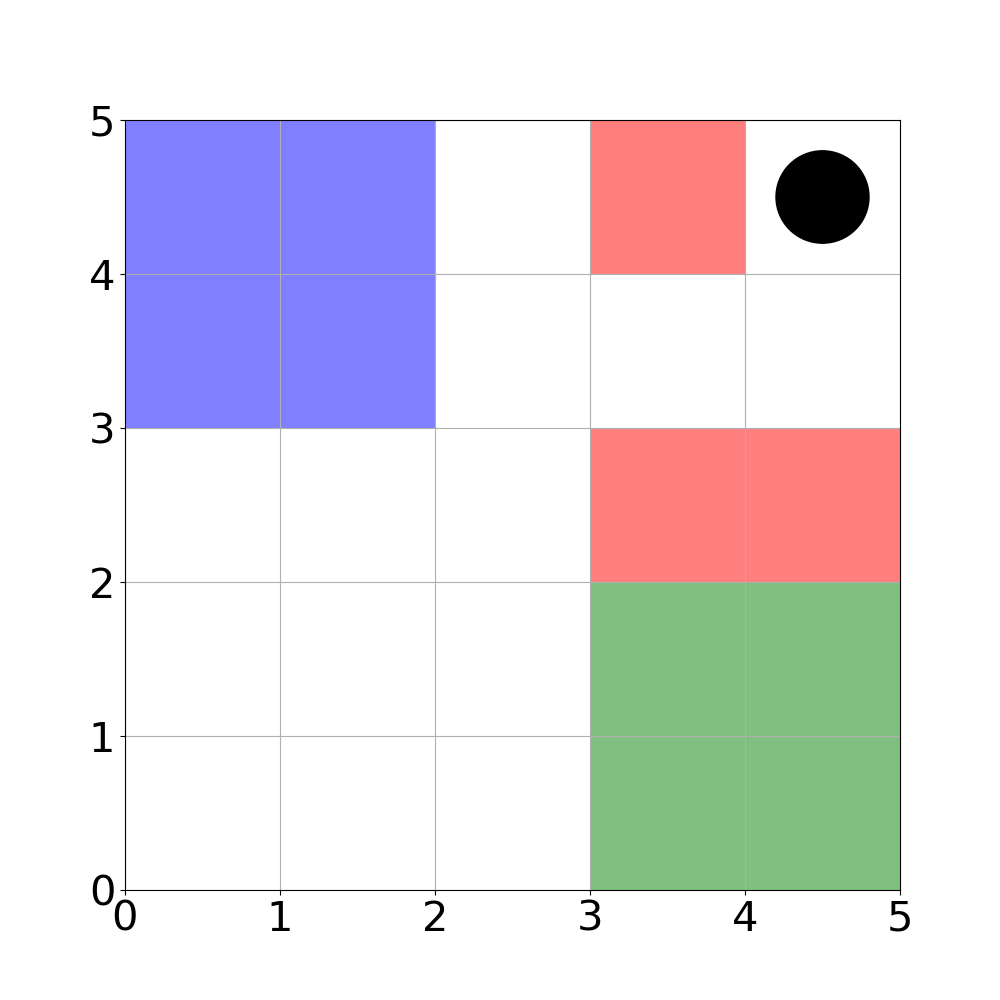}
        \label{fig:grid}
    }
     \subfigure[]{
        \centering
        \includegraphics[clip, width = 4.6cm]{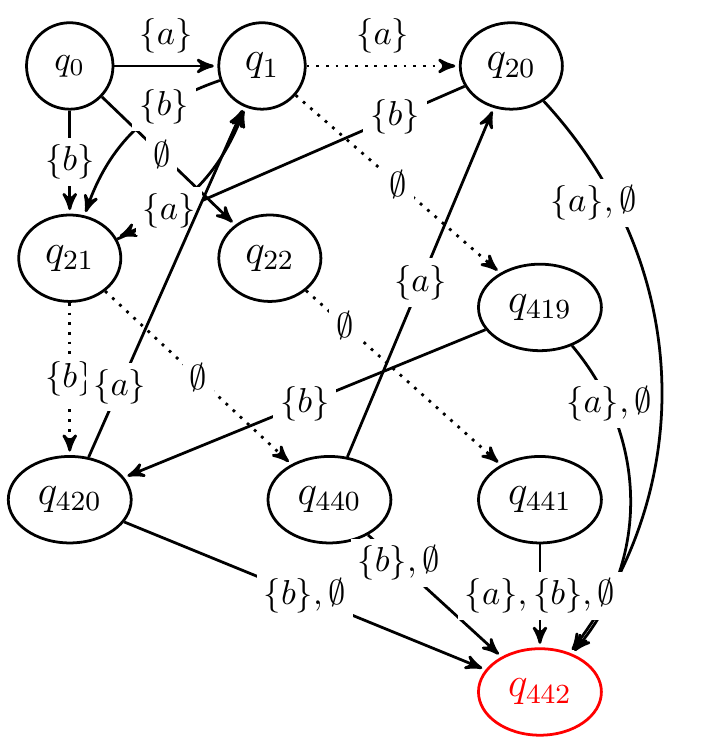}
        \label{fig:dKcBA}
      }
  \caption{(a) The grid world that the robot (black disk) moves. The blue and the green areas are the access point and the recharge point, respectively. The areas are labeled with $a$ and $b$, respectively. The red area is the dangerous area, which is labeled with $c$. (b) The d$K$cBA with $K=20$ converted from $\varphi_{ex}$. Each state is numbered with $q_i, i=0,\ldots,442$. For simplicity, we omit the states transitioned by the label $c$, other some states, and some edges. Each dotted line represents the consecutive transition by the label.}
\end{figure}
\end{runexample}
    
\section{Bounded Synthesis and Learning of Policy}
\label{Method}
We synthesize an optimal policy by the following 3 phases.
\begin{enumerate}
    \item We estimate $W$ and $W_p$ of $\mathcal{SM}^\otimes$. Simultaneously, we learn the dynamics $T$ and $D$ of $\mathcal{SM}$ within $W_p$ based on the Bayesian inference. After that, we learn the policy $\pi_{tr} : S^\otimes \setminus W \to A$ that achieves the maximum satisfaction probability of $\varphi_B$. Note that learning the behavior in and outside $W$ separately is based on the safety acceptance condition of d$K$cBA.
    \item Based on the value iteration, we synthesize a policy $\pi_{win} : W \to A$ that minimizes $V^{\mathsf{Risk}}$ utilizing the estimated $T$ and $D$. 
    \item We synthesize an optimal policy by combining $\pi_{tr}$ and $\pi_{win}$.
\end{enumerate}
We first describe phase 1). 
We define a reward function $\mathcal{R} : S^\otimes \to \mathbb{R}$ and a discount factor function $\Gamma : S^\otimes \to \{\gamma, \gamma_{acc} \}$ inspired by \cite{BWZP2020}, where $\gamma, \gamma_{acc} \in (0,1)$, as follows.
\begin{align}
    \mathcal{R} (s) =
    \left\{
    \begin{aligned}
      & (1 - \gamma_{acc})r_n & &\text{if} \ s \in Acc^\otimes , \\
      & 0 & &\text{otherwise},
    \end{aligned}
    \right.
    \label{def_reward}
  \end{align}
where $r_n$ is a negative value.

\begin{align}
      \Gamma (s) =
    \left\{
    \begin{aligned}
      & \gamma_{acc} & & \text{if}\ s\in Acc^{\otimes} \!,\\
      & \gamma & & \text{otherwise}.
    \end{aligned}
    \right.
    \label{def_discount}
  \end{align}
Then, we introduce the state-action value function $Q_{\pi} : S^\otimes \times A \to \mathbb{R}$ as follows. For any $s \in S$ and any $a \in A(s)$,
\begin{align*}
    Q_{\pi, \gamma}(s, a) = \mathbb{E}_{\pi}[\sum_{t=0}^{\infty} \mathcal{R}(s_{t+1})\prod_{k=0}^{t-1} \Gamma(s_{k+1})|s_0 = s, a_0 = a],
 \end{align*}
where $\mathbb{E}_{\pi}$ is the expected value under the policy $\pi$.
We call the following function $Q^\ast : S^\otimes \times A \to \mathbb{R}$ an \textit{optimal state-action value function}. For any $s \in S^\otimes$ and any $a \in A(s)$,
\begin{align}
    Q^\ast_\gamma(s,a) = \max_{\pi \in A^{S^\otimes}} Q_{\pi, \gamma}(s,a).
    \label{opt_Q}
\end{align}



\begin{algorithm}[t]
   \caption{Learning of Winning region and Dynamics}
   \label{LearnWD}
   \begin{algorithmic}[1]
   \renewcommand{\algorithmicrequire}{\textbf{Input:}}
   \renewcommand{\algorithmicensure}{\textbf{Output:}}
   \REQUIRE LTL formula $\varphi$, SMDP $\mathcal{SM}$, and $\ell$.
   \ENSURE Estimates of $W_p$, $h_T$, and $h_D$.
   \STATE Convert $\varphi$ to d$K$cBA $\det(B_{\varphi},K)$.
    \STATE Construct the product SMDP $\mathcal{SM}^{\otimes}$ of $\mathcal{SM}$ and $det(B_{\varphi},K)$.
    \STATE Initialize $Q:S^{\otimes} \times E^{\otimes} \to \mathbb{R}$ with -1 for any $(s,a) \in Acc^\otimes \times A$ and 0 otherwise.
    \STATE Initialize $W^0 = \{ s \in S^\otimes \;|\; \exists a \in A(s) \mbox{ s.t. } Q^0(s,a) = 0 \}$.
    \STATE Initialize $W_p^0 = \{ (s,a) \in S^\otimes \times A \;|\; Q^0(s,a) = 0 \}$.
    \STATE Initialize $\partial W^0 = \{ s \in W^0 | \exists (s,a) \in W^0_p, \exists s' \! \not \in W^0, \mbox{ s.t. } T(s'|s,a) \!>\! 0 \}$.
    \STATE Sample $s \in W^0$ and set $k=0$.
    \WHILE {$W^k_p$ does not converge.}
    \WHILE{True}
    \STATE Take action $a$ under $\pi^k_{ex}(s, \cdot)$. Observe $o = (s, a, s', \tau)$ and $r = \mathcal{R}(s,a,s')$. Append $o$ to $\mathcal{O}$.
    \IF{ $s' \not \in W^k$}
    \STATE Go to Line 16.
    \ENDIF
    \STATE $s \gets s'$.
    \ENDWHILE
    \STATE Remove all $(s, a, s', \tau)$ from $\mathcal{O}$ with respect to $(s,a)$.
    \STATE Update $Q^k$, $W^k$, $\partial W^k$, and $W^k_p$ using the update rules given by Function \ref{func_update}.
    \STATE Sample $s \in \partial W^k$.
    \STATE $k \gets k+1$.
    \IF{$k / \ell \in \mathbb{N}$}
    \STATE Calculate the posterior of $h_P$ and $h_D$ with observed data $\mathcal{O}$ within $W^k_p$.
    \ENDIF
    \ENDWHILE
   \end{algorithmic}
\end{algorithm}

We show the overall procedure to learn $W$, $W_p$, $T$, and $D$ of the SMDP in Algorithm \ref{LearnWD}. In Line 3, we initialize the state-action value $Q^0$. In Lines 4 and 5, we initialize $W^0$ and $W^0_p$ using $Q^0$ as the estimates of $W$ and $W_p$, respectively. In Line 6, we initialize $\partial W^0$ using $W^0$ and $W^o_p$, which is used in exploration. Next, in each episode $k$, we keep exploring in $W^k$ until leaving $W^k$. We explain the details of the exploration policy $\pi^k_{ex}$ later. During the learning, we observe dwell times $\tau \in \mathbb{R}_{\geq 0}$ as well as state transitions. In Line 16, we keep observed data only within $W_p^k$. We denote the set of observed data as $\mathcal{O} = \{ (s_i,a_i,s_{i+1},\tau_i) \}_{i \in I}$, where $I$ is an index set of the data. In Line 17, using Function \ref{LearnWWp}, $Q^k$ is updated and $W^k$, $W^k_p$, and $\partial W^k$ are updated by $Q^{k+1}$. Then, we sample a state from $\partial W^k$ and continue the learning.
We model $T$ as a parametric distribution $g_T(\cdot \;|\; {\theta^T}_{sa})$ for each $(s,a) \in S \times A$.
For each $(s,a,s') \in S \times A \times S$, we model $D(\cdot \;|\; s,a,s')$ as a parametric distribution $g_D(\cdot \;|\; \bm{\theta}^D_{sas'})$. We denote the sets of parameters of $g_T$ and $g_D$ as $\Theta^T$ and $\Theta^D$, respectively. In Line 20, we calculate their posteriors $h_T(\bm{\theta}^T_{sa} \;|\; \bm{\zeta}^T_{sa})$ and $h_D(\bm{\theta}^D_{sas'} \;|\; \bm{\zeta}^D_{sas'})$ based on (\ref{bayesrule}) every $\ell$ step, where $\ell \in \mathbb{N}$ is a given positive integer, with observed data.


\begin{algorithm}[t]
  \floatname{algorithm}{Function}
   \caption{Update Rules}
   \label{LearnWWp}
   \begin{algorithmic}[1]
   \renewcommand{\algorithmicrequire}{\textbf{Input:}}
   \renewcommand{\algorithmicensure}{\textbf{Output:}}
   \REQUIRE Transition $(s,a,s')$, reward $r$, and $Q^k$.
   \ENSURE $Q^{k+1}$, $W^{k+1}$, $\partial W^{k+1}$, $W^{k+1}_p$, and $\pi^k_{ex}$.
    \STATE $Q^{k+1}(s,a) \gets (1 - \alpha) Q^k(s,a) + \alpha \{ r + \max_{a'}Q^k(s',a') \}$.
    \STATE $W^{k+1} \gets \{ s \in S^\otimes \;|\; \exists a \in A(s) \mbox{ s.t. } Q^{k+1}(s,a) = 0 \}$.
    \STATE $W^{k+1}_p \gets \{ (s,a) \in S^\otimes \times A \;|\; Q^{k+1}(s,a) = 0 \}$.
    \STATE $\partial W^{k+1} \!=\! \{ s \in W^{k+1} | \exists (s,a) \in W^{k+1}_p, \exists s' \! \not \in W^{k+1}, \mbox{ s.t. } T(s'|s,a) \!>\! 0 \}.$
   \end{algorithmic}
   \label{func_update}
\end{algorithm}

\begin{assumption}
    The following conditions hold at phase 1.
 \begin{enumerate}
     \item The learning rate $\alpha$ is a constant.
     \item For each $s \in W^\infty$, $s$ is observed infinitely often w.p.1.
 \end{enumerate}
 \label{assum_W_s_infty}
\end{assumption}
It is known that $W^k$ (resp., $W^k_p$) is monotonically decreasing under Assumption \ref{assum_W_s_infty}, i.e., $W^k \subseteq W^{k'}$ (resp., $W^k_p \subseteq W^{k'}_p$) for any $k' \leq k$. Furthermore, $W^\infty := \lim_{k\to \infty} W^k = W$ (resp., $W^\infty_p := \lim_{k\to \infty} W^k_p = W_p$) holds w.p.1 \cite{OUS2021}.

\subsubsection*{Details of Exploration}

To facilitate the simultaneous collection of the transition and dwell time data efficiently, we use an entropy-based policy $\pi^k_{ent} : W^k_p \to [0,1]$ that satisfies the following conditions.
For any $k$ and any $(s^\otimes = (s,F), a), (s^\otimes,a') \in W^k_p$,
    \begin{align*}
     &\text{H}[h_T(\cdot \;|\; \bm{\zeta}^T_{s a})] + \frac{1}{n_{s a}}\sum_{s'}\text{H}[h_D(\cdot \;|\; \bm{\zeta}^D_{s a s'})]\\ 
     &\leq \text{H}[h_T(\cdot | \bm{\zeta}^T_{s a'})] + \frac{1}{n_{s a'}}\sum_{s'}\text{H}[h_D(\cdot \;|\; \bm{\zeta}^D_{s a' s'})]\\
     &\Rightarrow \pi^k_{ent}(s^\otimes,a) \leq \pi^k_{ent}(s^\otimes,a'),
    \end{align*}
    where $\text{H}[h(\cdot)] = -\int_x h(x) \ln{h(x)} dx$ is the entropy of $h$ and $n_{sa} = | \{ s' \in S \;|\; T(s' \;|\; s,a) > 0 \} |$.
Intuitively, $\pi^k_{ent}$ enhances to issue an action that triggers more informative transitions.
An estimate of $T$, denoted by $\tilde{T}$, is given as follows. For any $((s,F),a) \in W^k_p$ and any $s' \in S$,
\begin{align}
    \tilde{T}(s' \;|\; s,a) = \int g_T(s' \;|\; \bm{\theta}_{sa}^P ) h_T( \bm{\theta}_{sa}^T \;|\; \bm{\zeta}_{sa}^T) d\bm{\theta}_{sa}^T.
\end{align}
We define an estimate of the transition probability of the product SMDP $\tilde{T}^\otimes : S^\otimes \times S^\otimes \times A \to [0,1]$ for $\tilde{T}$ in accordance with (\ref{Ttimes}). 
To facilitate the efficient learning of $W_p$, we introduce $\pi^k_{W^\perp} : W^k_p \to [0,1]$ that satisfies the following condition. For any $k$ and any $(s^\otimes, a), (s^\otimes,a') \in W^k_p$,
 \begin{align*}
 &\sum_{s^{\otimes \prime} \not \in W^k} \tilde{T}^\otimes (s^{\otimes \prime} \;|\; s^{\otimes},a) \leq \sum_{s^{\otimes \prime} \not \in W^k} \tilde{T}^\otimes (s^{\otimes \prime} \;|\; s^{\otimes},a')\\
    &\Rightarrow \pi^k_{W^\perp}(s^{\otimes},a) \leq \pi^k_{W^\perp}(s^{\otimes},a').
 \end{align*}
Moreover, for each episode $k$, we restrict the region where an initial state is sampled to $\partial W^k$ given by
\[
\partial W^k \!=\! \{ s \in W^k | \exists (s,a) \in W^k_p, \exists s' \not \in W^k \mbox{ s.t. } T(s'|s,a) \!>\! 0 \}.
\]
Intuitively, $\pi^k_{W^\perp}$ and the state-sampling from $\partial W^k$ enhance to take an out-going transition from $W^k$ for each $k$.
We define an exploration policy $\pi^k_{ex} : W^k_p \to [0,1]$ as follows. For any $k$ and any $(s, a) \in W^k_p$, 
\begin{align}
    \pi^k_{ex}(s,a) = 
    \left\{
    \begin{aligned}
      & \pi^k_{W^\perp}(s,a) & &\text{if} \ s \in \partial W^k, \\
      & \pi^k_{ent}(s,a) & &\text{otherwise},
    \end{aligned}
    \right.
    \label{pi_ex}
\end{align}
$\pi^k_{ex}$ is expected to facilitate the efficient learning of $W_p$ and the unbiased learning of $T$ and $D$. 

Note that, for any transition $(s,a,s')$ stored in $\mathcal{O}$ after the learning, using $\pi^k_{ex}$, we have $(s,a) \in W_p$ and $s' \in W$ w.p.1 under Assumption \ref{assum_W_s_infty} \cite{OUS2021}.



After the learning of $W_p$, $T$, and $D$, we compute $\pi^\gamma_{tr} : S^\otimes \setminus W \to A$ defined as follows. For any $s \in S^\otimes \setminus W$,
\begin{align}
    \pi^\gamma_{tr}(s) \in \argmax_{a \in A(s)} Q^\ast_{\gamma}(s,a).
\end{align}

\begin{theorem}
\label{thmFW}
    Given a product SMDP $\mathcal{SM}^\otimes$ associated with an LTL formula $\varphi$, there exists a discount factor $\gamma' \in (0,1)$ such that, for any $\gamma > \gamma'$ and any $s \in S^\otimes$, the following equation holds. 
\begin{align}
 \label{Prast_Prmax_equi}
    Pr^{\mathcal{SM}^\otimes}_{\pi^\gamma_{tr}}(s \models \varphi_B) = \max_{\pi \in A^{S^\otimes}}Pr^{\mathcal{SM}^\otimes}_{\pi}(s \models \varphi_B).
\end{align}
\end{theorem}
A proof of Theorem 2 is similar to the proof of Theorem 3 in \cite{OUS2021} and omitted. From Theorem 2, for any $\gamma$ sufficiently close to one, $\pi^\gamma_{tr}$ maximizes the satisfaction probability of $\varphi_B$.

The learning scheme of $\pi^\gamma_{tr}$ is based on the update rule of Q-learning $Q(s,a) \gets (1 - \alpha_i) Q(s,a) + \alpha_i \{r + \max_{a'}Q(s',a')\}$, where $(s,a,s')$ and $r$ are the observed transition and the corresponding reward, respectively, and $i$ is the iteration number.
However, in the learning scheme, each learning episode is terminated when entering the estimated winning region $W^\infty$. We denote the learned $\pi^\gamma_{tr}$ as $\tilde{\pi}^\gamma_{tr}$. We assume the following condition.
\begin{assumption}
 $\sum_{i=1}^\infty \alpha_i = \infty$ and $\sum_{i=1}^\infty \alpha_i^2 < \infty$ hold.
 \label{assum_Qlearn}
\end{assumption}

\begin{figure}[htbp]
    \subfigure[]{
        \centering
        \includegraphics[clip, width = 4.1cm]{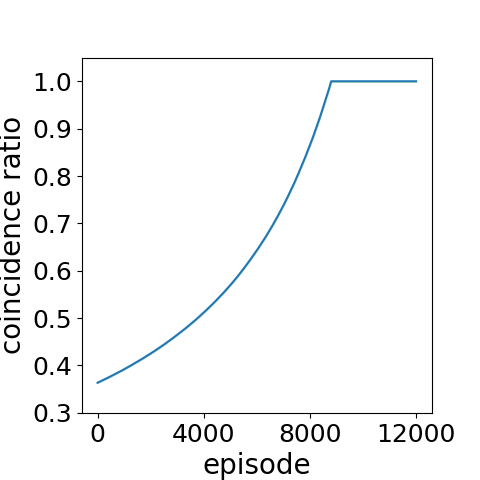}
        \label{fig:ave_Indk}
    }
    \subfigure[]{
        \centering
        \includegraphics[clip, width = 4.15cm]{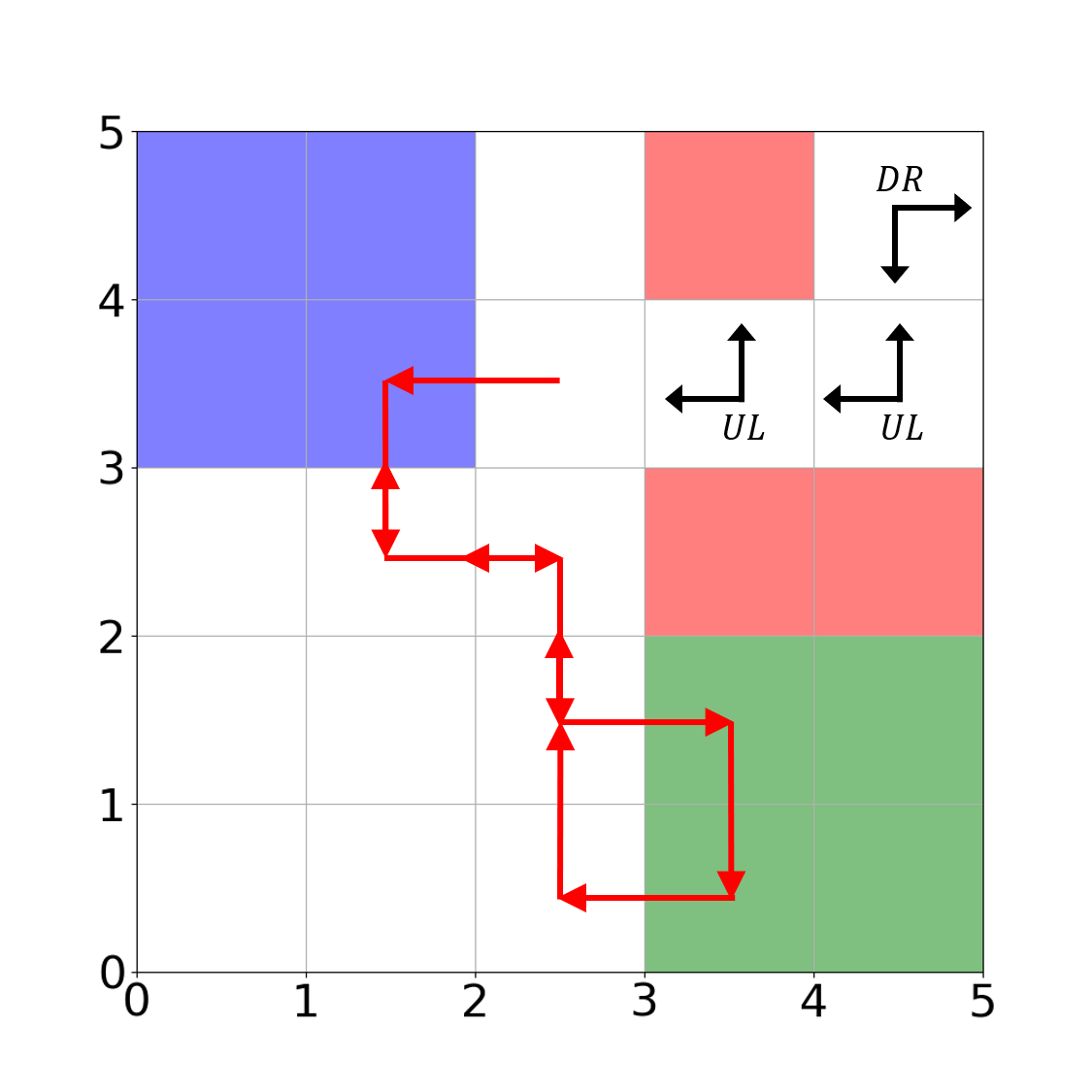}
        \label{fig:samplepath}
    }
    \caption{(a) The average of $Ind^k$ per episode. (b) The black arrows are illustrative actions taken by the learned robot at the states. The red arrows represents an obtained sample path.}
\end{figure}

To assess the convergence of $W_p^k$ for the running example, we introduce the index $Ind^k = |W_p|/|W_p^k|$ as the coincidence ratio between $W_p$ and $W_p^k$ for each $k \in \mathbb{N}_0$. We show the average of $Ind^k$ by repeating the learning 50 times, where the parameters are $\gamma = 0.9999$, $\gamma_{acc} = 0.9$, and $r_n = -1$. It converges around 9000 episode.
We show the action selection by $\tilde \pi_{tr}^\gamma$ at states of the SMDP corresponding to non-winning states as black arrows in Fig.\ \ref{fig:samplepath}. We iterate the Q-learning 5000 episodes to learn $\tilde \pi_{tr}^\gamma$. The actions imply that the robot moves to (3,4), which corresponds to a winning state, with the maximum probability.


Next, we describe phase 2). An estimate of $D$, denoted by $\tilde{D}$, is given as follows. For any $((s,F),a) \in W^\infty_p$ and any $s' \in S$,
\begin{align}
    &\tilde{D}(\tau | s,a,s') \!=\! \int g_D(\tau | \bm{\theta}_{sas'}^D ) h_D( \bm{\theta}_{sas'}^D | \bm{\zeta}_{sas'}^D) d\bm{\theta}_{sas'}^D.
\end{align}
We introduce $A_\varphi : W^\infty \to A$ as follows.
\begin{align}
  A_{\varphi}(s^\otimes) = \{ a \in A \;|\; (s^\otimes,a) \in W^\infty_p \}.
  \label{def_opt_SV}
\end{align}
Denoted by $\widetilde{\mathsf{Risk}}(s,a,s')$ is the risk of the transition $(s,a,s')$ for $\tilde{D}$, i.e., by (\ref{riskD}), we have $\widetilde{\mathsf{Risk}}(s,a,s') = f(\tilde{D}(s,a,s'))$ for any $(s,a,s') \in S \times A \times S$.
With $\tilde{T}$, $\widetilde{\mathsf{Risk}}$, and $A_\varphi$, we define $\tilde{Q}^{\mathsf{Risk}} : W^\infty_p \to \mathbb{R}$ as follows and compute it by the value iteration.
\begin{flalign*}
& \tilde{Q}^{\mathsf{Risk}}(s^\otimes, a) \\
&  = \sum_{s^{\otimes \prime} \in S^\otimes} \!\!\!\!\! \tilde{T}(s' | s,a) ( \widetilde{\mathsf{Risk}}(s,a,s') \!+\! \gamma_{\mathsf{r}} \min_{a' \in A_{\varphi}(s^\otimes)} \!\!\tilde{Q}^{\mathsf{Risk}}(s^{\otimes \prime},a')),
\end{flalign*}
where $s^\otimes = (s,F)$.
Then, we obtain $\tilde{\pi}_{win} : W^\infty \to A$ as follows.
\begin{align}
    \tilde{\pi}_{win}(s) \in \argmax_{a \in A_\varphi(s^\otimes)} \tilde{Q}^{\mathsf{Risk}}(s^\otimes,a).
    \label{def_piwin}
\end{align}
Finally, in phase 3), we obtain a policy $\pi^\infty_\gamma : S^\otimes \to A$ by combining $\pi^\gamma_{tr}$ and $\pi_{win}$. For any $s \in S^\otimes$, we define $\pi^\infty_\gamma$ as follows.
\begin{align}
    \tilde\pi^\ast_\gamma(s) = 
    \left\{
    \begin{aligned}
      &\tilde{\pi}_{win}(s) &&\text{ if } s \in W^\infty, \\
      &\tilde{\pi}^\gamma_{tr}(s) &&\text{ otherwise.}
    \end{aligned}
    \right.
    \label{def_piopt}
\end{align}

\begin{assumption}
 Assume the following conditions
 \begin{enumerate}
     \item For any $(s^\otimes \!=\! (s,F), a)\! \in \!W_p$ and any $(s',F') \in \{ s^{\otimes '} \in W^\infty \;|\; T(s^{\otimes '} | s^\otimes,a) > 0 \}$, there exist $\bm{\theta}^T_{sa}$ and $\bm{\theta}^D_{sas'}$ such that $T( \cdot \;|\; s,a) = g_T(\cdot \;|\; \bm{\theta}^T_{sa})$ and $D( \cdot \;|\; s,a,s') = g_D(\cdot \;|\; \bm{\theta}^D_{sas'})$.
     \item $g_T$ and $g_D$ satisfy Assumption \ref{assum_Doob}.
 \end{enumerate}
 \label{assum_gTD}
\end{assumption}
\begin{theorem}
    Let $\mathcal{SM}^\otimes$ be the product SMDP associated with an LTL formula $\varphi$. Under Assumptions \ref{assum_W_s_infty}, \ref{assum_Qlearn}, and \ref{assum_gTD}, there exist $\gamma' \in (0,1)$ and an optimal policy $\pi^\ast \in \Pi^\ast_{\varphi_B \succ \mathsf{Risk}}$ such that, for any $\gamma > \gamma'$, $\tilde\pi^\ast_\gamma = \pi^\ast$ in probability as $|\mathcal{O}| \to \infty$.
    \label{thm_optpol}
\end{theorem}
\begin{IEEEproof}
 Recall that $W^\infty_p = W_p$ and $W^\infty = W$ hold w.p.1 under Assumption \ref{assum_W_s_infty}. Thus, for any $(s,a) \in W_p$ and $s' \in \{ s' \in W \;|\; T(s' | s,a) > 0 \}$, we have $T(\cdot \;|\; s,a) = \tilde{T}(\cdot \;|\; s,a)$ and $D(\cdot \;|\; s,a,s') = \tilde{D}(\cdot \;|\; s,a,s')$ in probability as $|\mathcal{O}| \to \infty$ by Theorem \ref{th_Doob} under Assumptions \ref{assum_W_s_infty} and \ref{assum_gTD}. This implies $Q^{\mathsf{Risk}} = \tilde{Q}^{\mathsf{Risk}}$. Hence, for each $s \in W$, $\tilde{\pi}_{win}(s)$ maximizes $Q^{\mathsf{Risk}}(s, \cdot)$ on $A_\varphi(s)$ in probability. By Theorem \ref{thmFW} and the convergence property of Q-learning \cite{BT1996}, under Assumption \ref{assum_Qlearn}, there exists $\gamma'$ such that, for any $\gamma > \gamma'$, $\tilde{\pi}^\gamma_{tr}$ maximizes the satisfaction probability of $\varphi_B$ w.p.1. Therefore, there exists an optimal policy $\pi^\ast \in \Pi^\ast_{\varphi_B \succ \mathsf{Risk}}$ such that we have $\tilde\pi^\ast_\gamma = \pi^\ast$ in probability.
\end{IEEEproof}

By Theorem \ref{thm_optpol}, we can synthesize an optimal policy under the discount factor sufficiently close to 1 and the sufficient iteration in Algorithm \ref{LearnWD}.

In the running example, we model $T$ and $D$ as a categorical distribution and an exponential distribution, respectively. Prior for each distribution is chosen as a Dirichlet distribution and a Gamma distribution. We give the risk for transitions as $\mathsf{Risk}(s,a,s') = \mu_{sas'} + \sigma_{sas'} $, where $\mu_{sas'}$ and $\sigma_{sas'}$ are the mean and the standard deviation of the estimated dwell time. We set the parameters as $\gamma_{\mathsf{r}} = 0.9$ and $\ell=1$. We show a sample path by the learned $\tilde \pi_\gamma^\ast$ as the red arrows in Fig.\ \ref{fig:samplepath}. We observe that the robot tends to move near (3,3) under the satisfaction of $\varphi_{ex}$ after learning, which implies that the robot minimizes the long-term risk while satisfying $\varphi_{ex}$.

\section{Conclusion}
In this letter, for a semi-Markov decision process, we proposed a learning-based bounded synthesis of the optimal policy that minimizes a long-term risk for the dwell time under the satisfaction of a given LTL specification. We show that the synthesized policy is equal to an optimal one in probability under some assumptions and conditions. Future works are to extend the proposed method to the continuous state space and to develop the learning method to estimate the winning region and the policy to minimize an accumulative risk concurrently.
It is also future work to apply the proposed method to a real system.

\bibliographystyle{ieeetr}
\bibliography{reference}
\end{document}